\documentclass[twocolumn,twoside,slac_two]{revtex4}
\usepackage{fancyhdr}
\usepackage{graphicx}
\newcommand{\apj}{ApJ}
\newcommand{\apjl}{ApJL}
\newcommand{\apjs}{ApJS}
\newcommand{\aap}{A\&A}
\newcommand{\aaps}{A\&AS}
\newcommand{\mnras}{MNRAS}

\begin{document}

\title{Discovery and identification of two gamma-ray blazars at
low galactic latitude with VERITAS}

%

\author{M. Errando}
\affiliation{Department of Physics \& Astronomy, Barnard College, Columbia University, 3009 Broadway, New York, NY 10027, USA}
\author{for the VERITAS Collaboration}

\begin{abstract}
We report on the discovery of two gamma-ray sources at $E>200$ GeV with VERITAS: RX~J0648.7+1516 and VER~J0521+211. Both sources are located at low galactic latitudes ($|b| < 10^\circ$), and were previously classified as active galactic nuclei of unknown type. The discovery and characterization of gamma-ray emission together with follow-up optical spectroscopy permitted the identification of RX~J0648.7+1516 and VER~J0521+211 as BL Lac-type blazars. In the case of RX~J0648.7+1516, a spectroscopic redshift of $z=0.179$ was derived.
\end{abstract}

\maketitle

\thispagestyle{fancy}

\section{Introduction}
Blazars are active galactic nuclei (AGN) hosting supermassive black holes, with relativistic jets pointing along the line of sight to the observer. The small viewing angle of the jet makes it possible to observe strong relativistic effects, such as a boosting of the emitted power and a shortening of the characteristic time scales. Blazars are typically identified through radio and optical surveys, being radio loudness, flat spectrum and compact morphology in the radio band, some degree of optical polarization, and fast, large amplitude variability their defining observational characteristics. Close to the Galactic plane ($|b|\lesssim10^\circ$), diffuse radio emission and confusion with local radio sources make candidate blazars difficult to select. Heavy optical extinction due to interstellar dust \cite{schlegel} further complicates spectroscopical identification of blazars at low latitudes. These observational challenges give us only a fragmentary knowledge of the blazar population in the galactic plane a region. Several blazar catalogs directly avoid galactic latitudes in order not to deal with the inevitable incompleteness of radio and optical surveys in the region \cite[e.g.,][]{crates,cgrabs,bzcat}.

\begin{figure}[b]
\centering
\includegraphics[width=8.5cm]{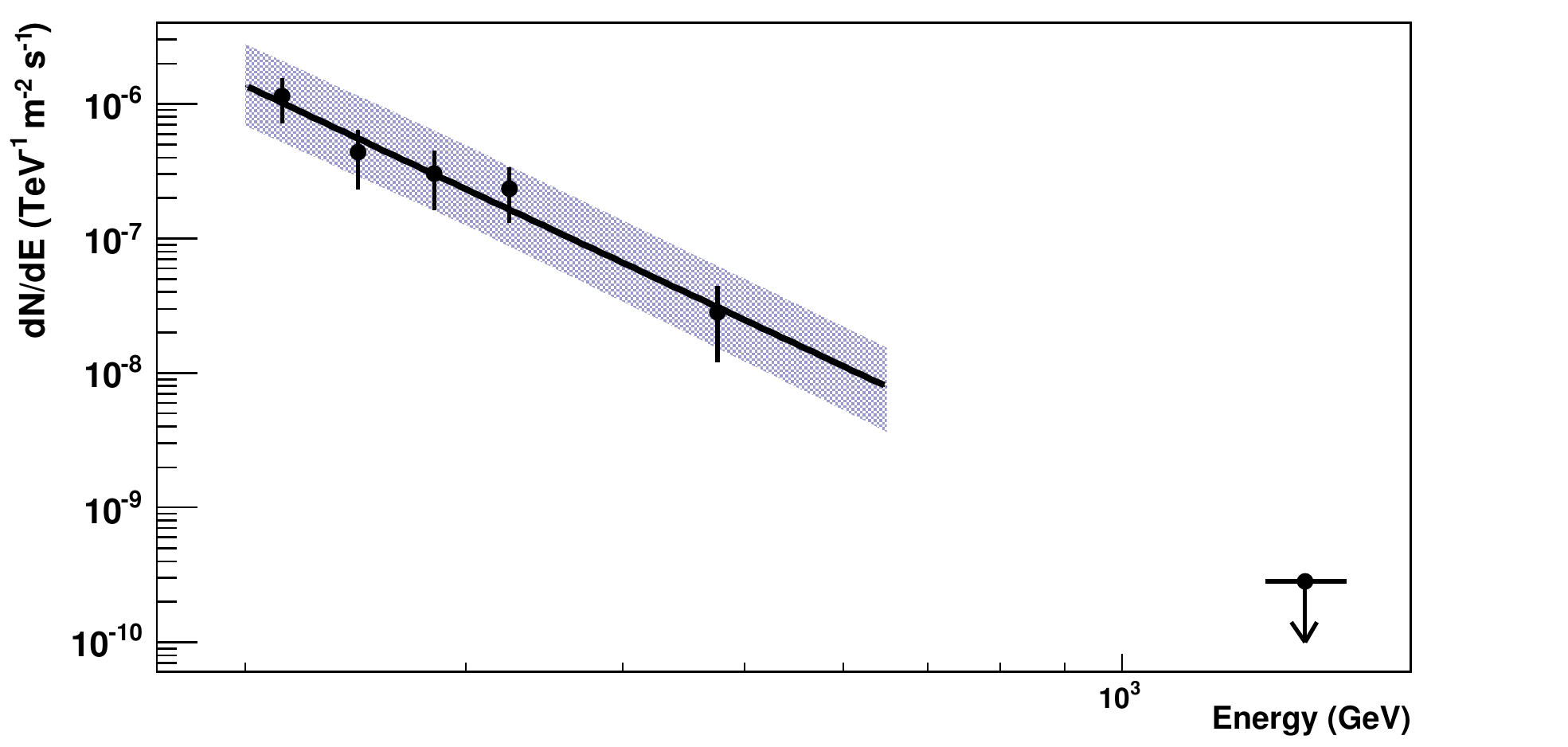}
\caption{The differential photon spectrum of RX J0648.7+1516 between 200 and 650 GeV measured by VERITAS between 4 March and 15 April 2010 (MJD 55259-55301). The solid line shows a power-law fit to the measured flux derived with four equally log-spaced bins and a final bin boundary at 650 GeV, above which there are few on-source photons. A 99\% confidence upper limit evaluated between 650 GeV and 5 TeV assuming a photon index of 4.4 is also shown. The shaded region shows the systematic uncertainty of the fit, which is dominated by 20\% uncertainty on the energy scale.} \label{fig:0648-gamma}
\end{figure}

A fraction of blazars are detected in the gamma-ray band. The second catalog of {\em Fermi}-detected AGNs \cite[2LAC, ][]{2lac} identifies 861 blazars in the energy band between 100 MeV and 300 GeV. In the very high energy range (VHE, $E>100$ GeV), ground-based imaging atmospheric Cherenkov telescopes (IACTs) have detected 43 blazars so far \cite{tevcat}. Gamma-rays do not suffer from significant absorption at low galactic latitudes. Even if the gamma-ray diffuse emission is brighter close to the galactic plane, observations in the gamma-ray band have shown to be a successful tool to identify blazars located behind the Galactic plane, which are heavily absorbed at other wavelengths. Two low-latitude blazars were discovered by association with EGRET sources \cite{muk,halpern,sguera}. More recently, associations with unidentified {\em Fermi}-LAT sources have led to the identification of new blazars located behind the galactic plane \cite{bassani,mirabal,vanden}. Identifications, blazar counterparts are found for 102 {\em Fermi}-LAT sources at $|b|\lesssim10^\circ$ in 2LAC. In the VHE band, MAGIC~J2001+435 \cite{mariotti} and HESS~J1943+213 \cite{hess} are recent examples of the capability of ground-based IACTs to reveal new blazars shining through the galactic plane by their gamma-ray emission.

\begin{figure*}[t]
\centering
\includegraphics[width=0.48\textwidth,angle=90]{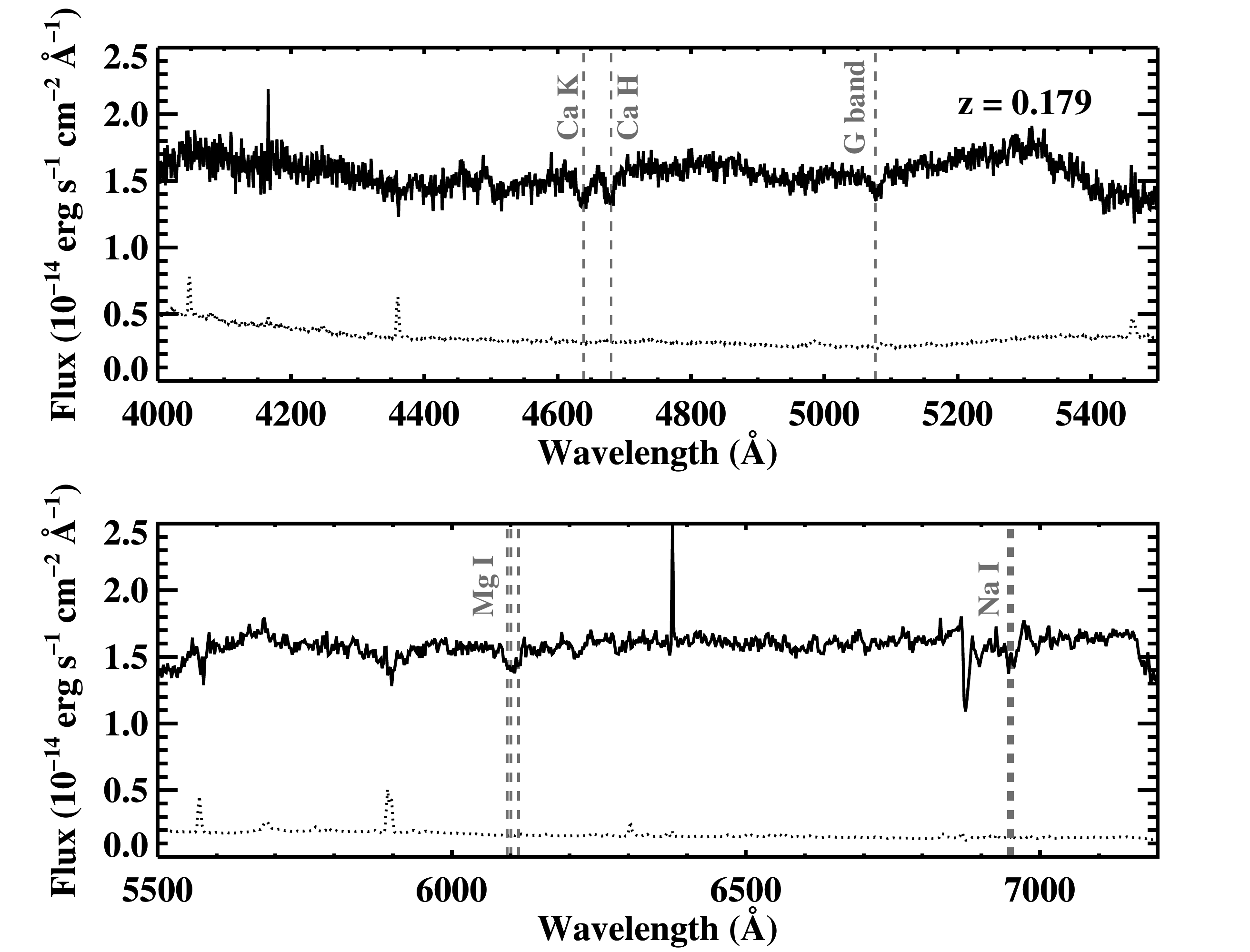}
\caption{Optical spectrum of RX~J0648.7+1516 showing the Ca H+K, G-band, Na I and Mg I spectral features, indicating a redshift of $z = 0.179$. The blazar was observed at Lick Observatory using the 3-meter Shane Telescope.} \label{fig:0648-opt}
\end{figure*}

This proceedings report the VHE detection of two low-latitude gamma-ray-emitting blazars with VERITAS: RX~J0648.7+1516 and VER~J0521+211. VERITAS \cite{veritas} is an array of four imaging atmospheric Cherenkov telescopes located at the Fred Lawrence Whipple Observatory (FLWO) in southern Arizona (31 40\,N, 110 57\,W,  1.3\,km a.s.l.). It combines a large effective area over a wide range of energies (100\,GeV to 30\,TeV) with an energy resolution of 15-25\% and an angular resolution of less that $0.1^{\circ}$. The high sensitivity of VERITAS allows the detection of sources with a flux of 0.01 times that of the Crab Nebula in about 25 hours. The standard VERITAS analysis methods are described in \cite{cogan, daniel}. A more detailed description of the gamma-ray detections of RX~J0648.7+1516 and VER~J0521+211 together with a complete multi wavelength characterization can be found in \cite[][respectively]{0648,0521}.

\section{RX~J0648.7+1516}
RX~J0648.7+1516 is a ROSAT-detected X-ray source associated with a compact flat-spectrum radio source \cite{rgb}, located $6.3^\circ$ off of the galactic plane. Having the typical characteristics of a radio-loud AGN, no definitive optical identification was found in the literature, with several unsuccessful attempts to find an optical counterpart \cite{brink,motch,haako}. {\em Fermi}-LAT discovered a gamma-ray source associated with RX~J0648.7+1516 \cite{1fgl}. The source was identified as a promising VHE candidate by the {\em Fermi}-LAT collaboration, and this information triggered VERITAS observations of the source.

RX~J0648.7+1516 was observed with VERITAS between 4 March and 15 April 2010 (MJD 55259-55301), for a total exposure of 19.3 hours after rejecting data taking under poor weather conditions or with hardware problems. After event parameterization and background rejection, an excess of 283 gamma-ray-like events from the direction of RX~J0648.7+1516 was found, corresponding to a significance of $5.2\sigma$. Figure~\ref{fig:0648-gamma} shows the differential energy spectrum of the obtained signal, which can be parametrized by a power law of the form $\mathrm{d}N/\mathrm{d}E = N_0 \times (E/0.3\, \mathrm{TeV})^{-\Gamma}$ with $N_0=(2.3\pm0.5)\times 10^{-11}$ cm$^{-2}$s$^{-1}$ and $\Gamma = 4.4\pm0.8$. The excess event distribution of the signal is consistent with a point-like source and compatible within errors with the radio position of RX~J0648.7+1516. A VHE light curve in 1-day bins does not show significant variability during VERITAS observations.

The detection in the VHE band strongly suggests the source being a blazar, since most of the VHE-detected AGNs are of the blazar class. To test that hypothesis, optical spectroscopy measurements were triggered after the VERITAS detection. Two spectra were obtained with the KAST double spectrograph on the 3-meter Shane telescope at the UCO/Lick Observatory (see Figure~\ref{fig:0648-opt}). The optical spectrum shows Ca H+K, G-band, Na I and Mg I  spectral absorption features, compatible with the source being at a redshift of $z = 0.179$.  These observations also provide a definitive optical identification of RX~J0648.7+1516 as a blazar. The fact that the absorption lines have equivalent widths $<5 \mathrm{\AA}$ indicate a blazar of the BL Lac sub-class. 

\begin{figure*}[t]
\centering
\includegraphics[width=0.7\textwidth]{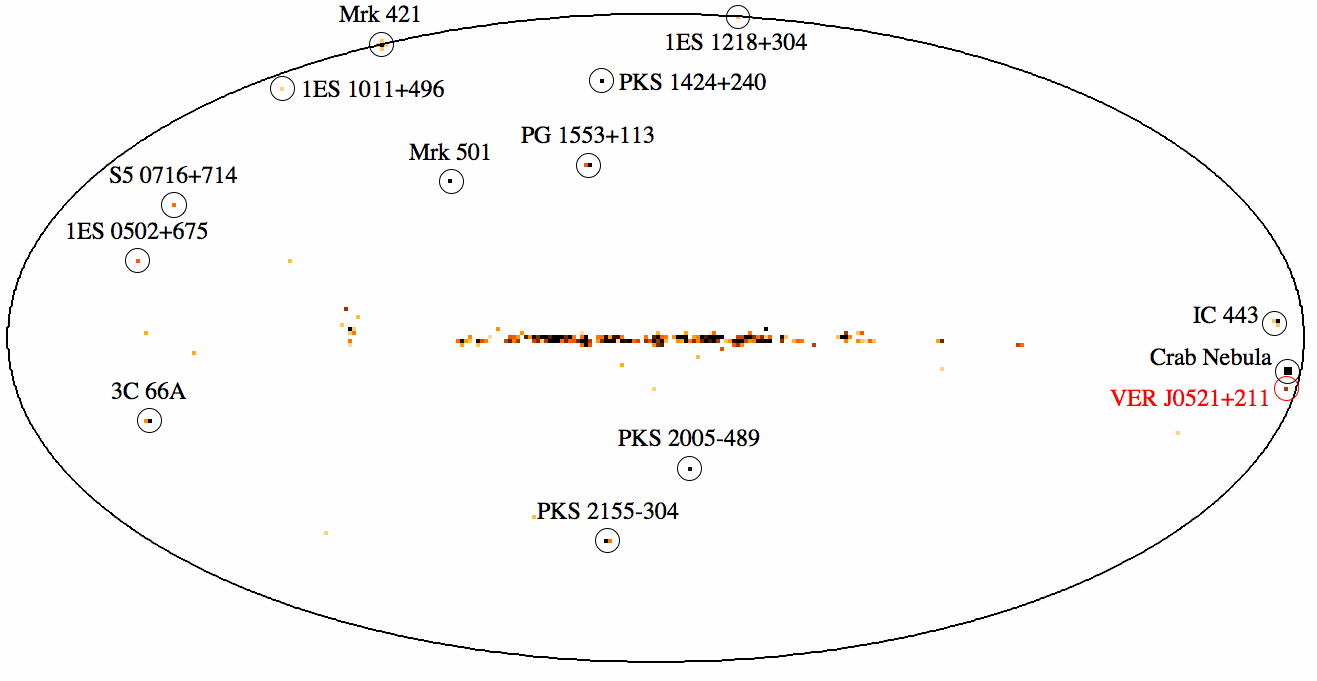}
\caption{All-sky counts map for the 1st year of Fermi-LAT data at energies above 50\,GeV. Most of the hotspots in the map correspond to known TeV sources. VERITAS observations of hotspots not associated with VHE sources led to the detection of VER J0521+211 (in red).}
\label{fig:map}
\end{figure*}

\section{VER~J0521+211}
After the public release of the 1st year of data accumulated by {\em Fermi}, an analysis of the photons with highest energies collected by {\em Fermi}-LAT was carried out inside the VERITAS collaboration in order to look for VHE gamma-ray candidates. In particular, photons with $E>50$\,GeV recorded by {\em Fermi}-LAT during its first year of operations were binned in an all-sky map (Figure~\ref{fig:map}). Most of the hotspots in the map could be associated with known VHE sources, proving the ability of the technique to select good candidates for VHE observations. One of the few previously undetected hotspots was less than $0.1^\circ$ away from RGB~J0521.8+2112: a radio-loud AGN with significant X-ray emission located $8.7^\circ$ off of the galactic plane. VERITAS observations centered at the position of RGB~J0521.8+2112 were taken between 22 and 24 October 2009 (MJD 55126-55128), leading to the detection of a new VHE gamma-ray source: VER~J0521+211 \cite{0521-atel1}. An excess with significance of $5.5\sigma$ was obtained during the first 230 minutes of exposure. VERITAS continued monitoring the new gamma-ray source and detected a high flux state in 22 November 2009 (MJD 55157), when VER~J0521+211 reached a peak flux $>3$ times brighter than the discovery observations \cite{0521-atel2}.

\begin{figure*}[t]
\centering
\includegraphics[height=2.2in]{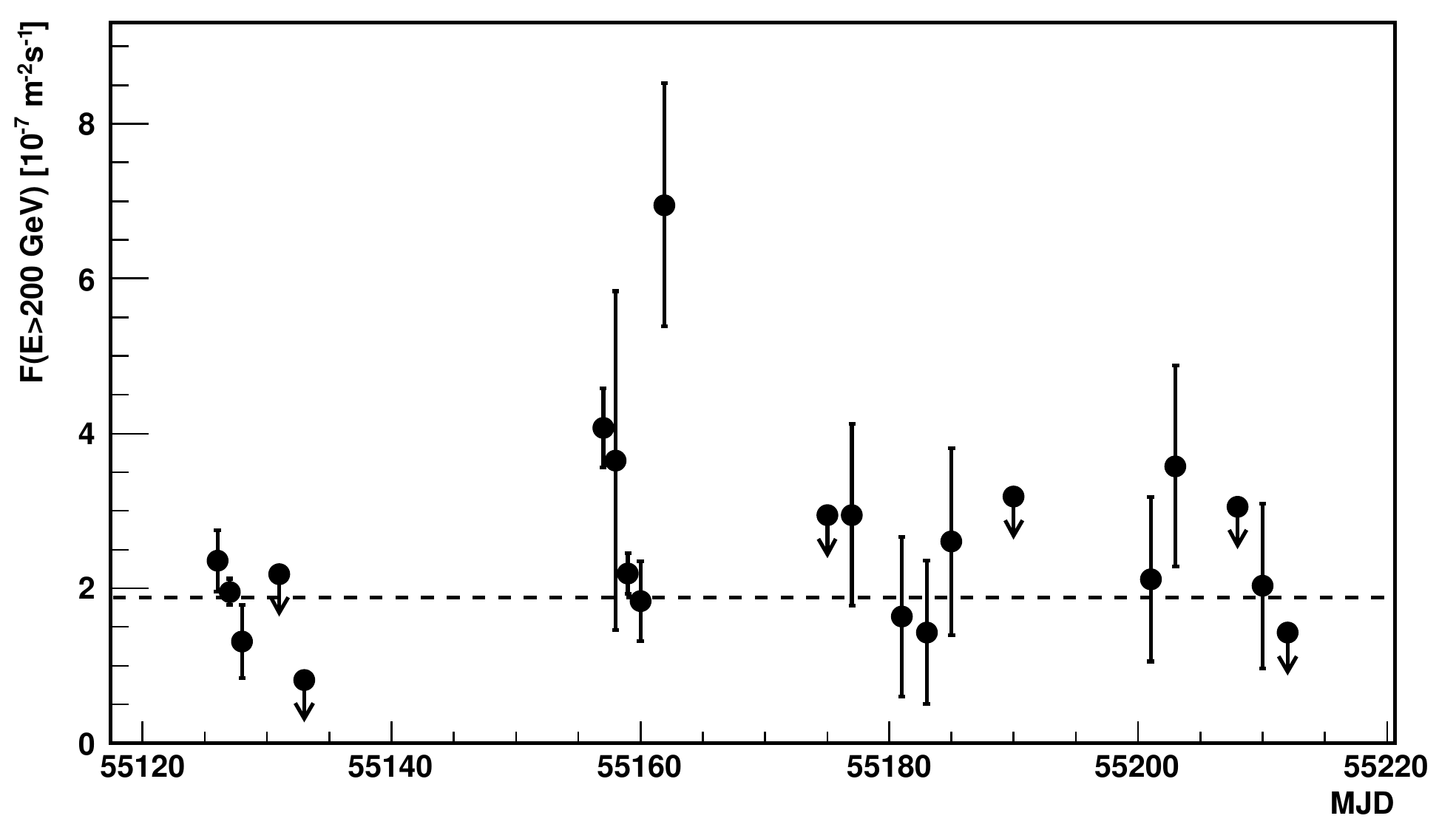}
\includegraphics[height=2.2in]{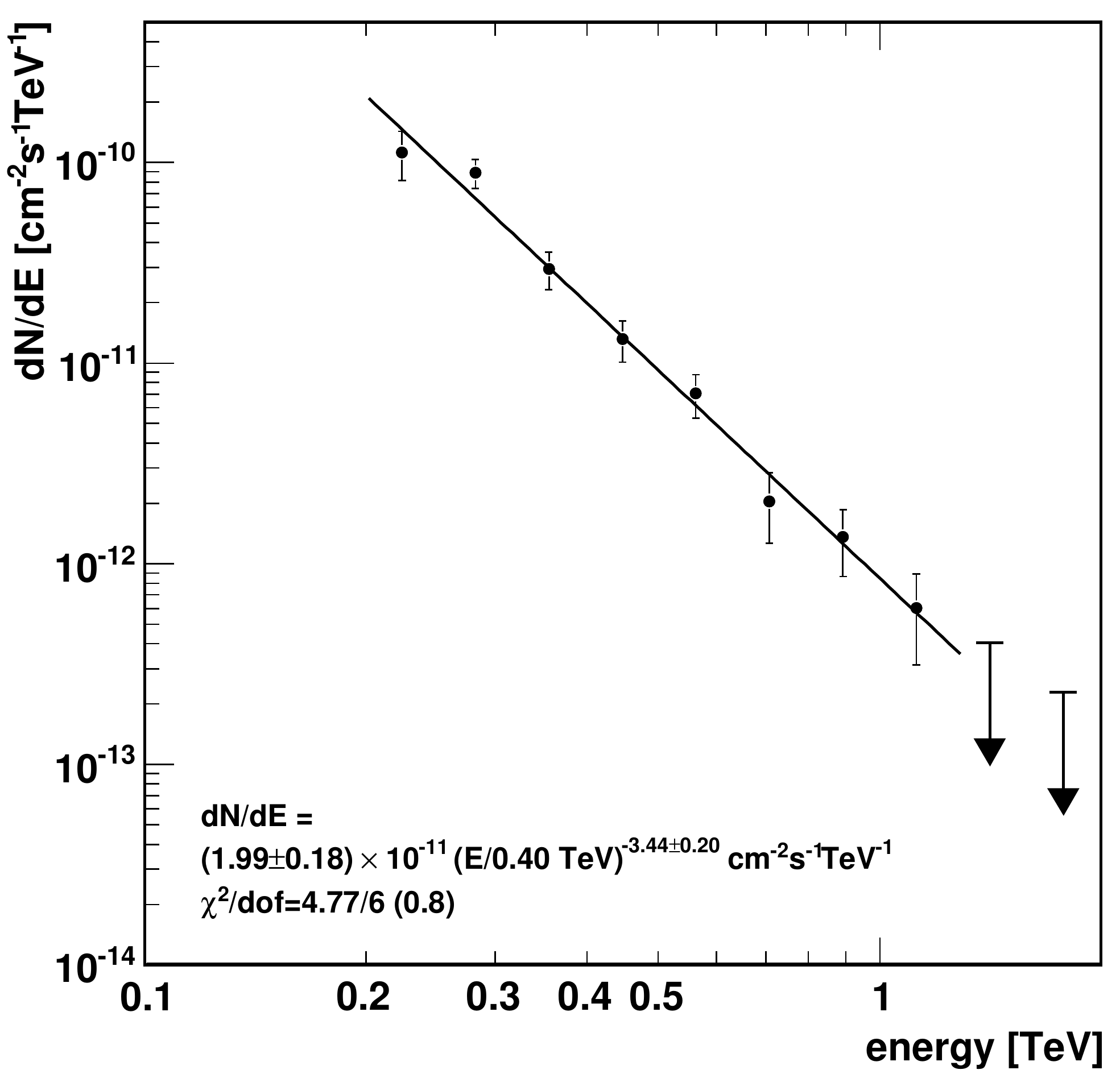}
\caption{{\em Left:} Daily gamma-ray light curve of VER~J0521+211 at $E>200$ GeV as measured by VERITAS. Significant variability is seen, with a major flare in November 2010. {\em Right:} Gamma-ray spectrum of VER~J0521+211 as measured with VERITAS between 0.2 and 1 TeV.} \label{fig:0521}
\end{figure*}

A total exposure of 14.5 hours was selected integrating all the data accumulated between 22 October 2009 and 16 January 2010 (MJD 55126-55212), after rejecting data taken under poor weather conditions, or with hardware problems. A signal with significance of $15.6\sigma$ was obtained. The excess distribution of VER~J0521+211 is consistent with a point-like source and its location compatible with the radio position of RGB~J0521.8+2112. Figure~\ref{fig:0521} shows the derived gamma-ray light curve for energies $>200$ GeV in 1-day time bins and the energy spectrum obtained between 200 GeV and 1 TeV. The light curve shows significant variability. The differential energy spectrum is well described by a power-law function of the form $\mathrm{d}N/\mathrm{d}E = N_0 \times (E/0.4\, \mathrm{TeV})^{-\Gamma}$ with $N_0=(1.99\pm0.18)\times 10^{-11}$ cm$^{-2}$s$^{-1}$ and $\Gamma = 3.44\pm0.20$.

There was no obvious affiliation for RGB~J0521.8+2112 prior to the VERITAS detection, other than RGB sources being generally radio-loud AGNs \cite{rgb}. Gamma-ray loudness in AGNs is typically interpreted as doppler-boosted emission from a relativistic jet closely oriented towards the line of sight to the observer. Therefore, the VHE detection of VER~J0521+211 strongly suggests a blazar-type AGN. Most VHE-detected AGN are blazars (43 out of 47), with the only exceptions being nearby FR\,I radio galaxies. Following the VHE detection, optical spectroscopy observations where performed at the 6.5-meter MMT telescope at the FLWO. The observations revealed a continuum-dominated spectrum, identifying the optical counterpart as a BL Lac-type blazar.  No spectral absorption lines could be identified, and therefore the redshift of VER~J0521+211 remains unknown.

\section{Discussion and conclusions}
VERITAS observed and detected gamma-ray emission from two previously unidentified radio-loud AGNs: RX~J0648.7+1516 and VER~J0521+211. The detection in the VHE band together with follow-up optical spectroscopy allowed to identify both newly discovered sources as gamma-ray-emitting blazars of the BL Lac sub-class. In the case of RX~J0648.7+1516, a redshift of $z=0.179$ could be derived from the optical observations. 

Due to their location close to the galactic plane, these two relatively powerful gamma-ray blazars were not classified as such before the VERITAS detection, and where not in the lists of candidate targets for VHE observations \cite[e.g.,][]{donato,costamante}. The successful identification of RX~J0648.7+1516 and VER~J0521+211 through their gamma-ray emission suggests that VHE observations
can be a useful tool to identify blazars at low galactic latitudes, where classical blazar surveys conducted in the radio and optical bands suffer from source confusion and optical extinction. 

\bigskip 
\begin{acknowledgments}
This work was supported in part by the NSF grant Phy-0855627 and NASA grant
NNX10AP66G. VERITAS research is supported by grants from the US Department
of Energy, the US National Science Foundation, and the Smithsonian Institution, by
NSERC in Canada, by Science Foundation Ireland, and by STFC in the UK. We
acknowledge the excellent work of the technical support staff at the FLWO and at the
collaborating institutions in the construction and operation of the instrument.
\end{acknowledgments}

\bigskip 

\end{document}